\documentclass{llncs}

\usepackage{amsmath, amssymb}
\usepackage{bussproofs}
\usepackage{stmaryrd}

\newcommand{\ignore}[1]{}
\newcommand{\ital}[1]{{\em #1}}

\newcommand{\eg}{{\rm e.g.}}
\newcommand{\ra}{\rightarrow}

\newcommand{\lprolog}{$\lambda$Prolog}
\newcommand{\hhf}{$hohh$}
\newcommand{\lftype}{\text{\ital{lf-type}}}
\newcommand{\lfobject}{\text{\ital{lf-obj}}}

\newcommand{\subs}[2]{#2[#1]}
\newcommand{\invup}[4]{inv^{\uparrow}(#1;#2;#3)=#4}
\newcommand{\invdown}[4]{inv^{\downarrow}(#1;#2;#3)=#4}
\newcommand{\lfprovesig}[3]{#1 \, \vdash_{#2} \, #3}
\newcommand{\invAbs}{\mbox{\sl inv-abs}}
\newcommand{\invApp}{\mbox{\sl inv-app}}
\newcommand{\invConst}{\mbox{\sl inv-const}}
\newcommand{\invSyn}{\mbox{\sl inv-syn}}
\newcommand{\invVar}{\mbox{\sl inv-var}}

\newcommand{\ruvappt}{{\scriptsize APP$_\text{t}$}}
\newcommand{\ruvpit}{{\scriptsize PI$_\text{t}$}}
\newcommand{\ruvctxt}{{\scriptsize CTX$_\text{t}$}}
\newcommand{\ruvinito}{{\scriptsize INIT$_\text{o}$}}
\newcommand{\ruvappo}{{\scriptsize APP$_\text{o}$}}
\newcommand{\ruvabso}{{\scriptsize ABS$_\text{o}$}}

\newcommand{\myvec}[1]{\overrightarrow{#1}}

\newcommand{\subst}[2]{#1[#2]}

\newcommand{\hastype}[2]{hastype \ #1 \ #2}

\newcommand{\encTerm}[1]{\langle #1 \rangle}
\newcommand{\enc}[1]{\lbrace\!\!\lbrace #1 \rbrace\!\!\rbrace}
\newcommand{\encExtP}[2]{\llbracket #1 \rrbracket^{+}_{#2}}
\newcommand{\encExtN}[1]{\llbracket #1 \rrbracket^{-}}
\newcommand{\termUV}[2]{#1 \sqsubset_o #2}
\newcommand{\formulaUV}[2]{#1 \sqsubset_t #2}

\newcommand{\forallx}[2]{\forall #1.#2}

\newcommand{\sigsequent}[3]{#1 ; #2 \longrightarrow #3}

\newcommand{\lambdax}[2]{\lambda #1.#2}
\newcommand{\typedlambda}[3]{\lambda #1 \mbox{:} #2 . #3}

\newcommand{\typedpi}[3]{\Pi #1 \mbox{:} #2 . #3}

\newcommand{\app}[2]{#1\ #2}
\newcommand{\emptyctx}{\cdot}
\newcommand{\oftype}[2]{#1 : #2}
\newcommand{\ctx}{\mbox{\sl ctx}}
\newcommand{\sig}{\mbox{\sl sig}}
\newcommand{\kind}{\mbox{\sl kind}}
\newcommand{\type}{\mbox{\sl Type}}

\newcommand{\bcGoal}{\mbox{\sl backchain}}
\newcommand{\topGoal}{$\top \mbox{\sl R}$}

\newcommand{\impGoal}{$\supset\! \mbox{\sl R}$}
\newcommand{\allGoal}{$\forall \mbox{\sl R}$}

\newcommand{\varext}{\mbox{\sl meta-var}}

\newcommand{\varobj}{\mbox{\sl var-obj}}
\newcommand{\constobj}{\mbox{\sl const-obj}}

\newcommand{\appobj}{\mbox{\sl app-obj}}
\newcommand{\absobj}{\mbox{\sl abs-obj}}

\newcommand{\ie}{{\rm i.e.}}
\newcommand{\etal}{{\rm et. al.\ }}

\newtheorem{theorem'}[theorem]{Theorem}
\newtheorem{corollary'}{Corollary}
\newtheorem{lemma'}[theorem]{Lemma}

\title{A $\lambda$Prolog Based Animation\\ of Twelf Specifications}
\author{Mary Southern \and Gopalan Nadathur}
\institute{University of Minnesota, Minneapolis MN 55455, USA}

\begin{document}
  \maketitle

  \begin{abstract}
  \label{sec:abstract}
Specifications in the Twelf system are based on a logic programming
interpretation of the Edinburgh Logical Framework or LF.
We consider an approach to animating such specifications using a
$\lambda$Prolog implementation.
This approach is based on a lossy translation of the dependently typed
LF expressions into the simply typed lambda calculus (STLC) terms of
$\lambda$Prolog and a subsequent encoding of lost dependency
information in predicates that are defined by suitable clauses.
To use this idea in an implementation of logic programming {\it a la}
Twelf, it is also necessary to translate the results found for
$\lambda$Prolog queries back into LF expressions.
We describe such an inverse translation and show that it has the
necessary properties to facilitate an emulation of Twelf behavior
through our translation of LF specifications into $\lambda$Prolog
programs.
A characteristic of Twelf is that it permits queries to consist of
types which have unspecified parts represented by meta-variables for
which values are to be found through computation.
We show that this capability can be supported within our translation
based approach to animating Twelf specifications.

  \end{abstract}
  
  \section{Introduction}
\label{sec:intro}
The Edinburgh Logical Framework or LF \cite{harper93jacm} is a
dependently typed lambda calculus that has proven useful
in specifying formal systems such as logics and programming
languages (see, \eg, \cite{harper07jfp}).
The key to its successful application in this setting is twofold.
First, the abstraction operator that is part of the syntax of LF
provides a means for succinctly encoding formal objects whose
structures embody binding notions.
Second, LF types can be indexed by terms and, as such, they can be
used to represent relations between objects that are encoded by terms.
More precisely, types can be viewed as formulas and type checking as a
means for determining if a given term represents a proof of that
formula.
Proof search can be introduced into this context by interpreting
a type as a request to determine if there is a term of that
type.
Further, parts of a type can be left unspecified, thinking of it then
as a request to fill in these parts in such a way that the resulting
type is inhabited.
Interpreting types in this way amounts to giving LF a logic
programming interpretation.
The Twelf system \cite{pfenning91lf,pfenning02guide}
is a realization of LF that is based on such an interpretation.

An alternative approach to specifying formal systems is to use a
predicate logic.
Objects treated by the formal systems can be represented by the terms
of this logic and relations between them can be expressed through
predicates over these terms.
If the terms include a notion of abstraction, \eg, if they encompass
simply typed lambda terms, then they provide a convenient means for
representing binding notions.
By restricting the formulas that are used to model relations suitably,
it is possible to constrain proof search behavior so that the formulas
can be given a rule-based interpretation.
The logic of higher-order hereditary Harrop formulas (\hhf) has been
designed with these ideas in mind and many experiments have shown this
logic to be a useful specification device (see, \eg,
\cite{miller12proghol}).
This logic has also been given a computational interpretation in the
language \lprolog~\cite{nadathur88iclp}, for which efficient
implementations such as the Prolog/Mali~\cite{brisset92tr} and the
Teyjus~\cite{teyjus.website} systems have been developed.

The two different approaches to specification that are described above
have a relationship that has been explored formally.
In early work, Felty and Miller showed that LF derivations could be encoded
in \hhf~derivations by describing a translation from the former to the
latter \cite{felty90cade}.
This translation demonstrated the expressive power of \hhf, but did
not show the correspondence in proof search behavior.
To rectify this situation, Snow~\etal described a transformation of LF
specifications into \hhf~formulas that allowed the construction of
derivations to be related~\cite{snow10ppdp}.
This work also showed how to make the translation more efficient
by utilizing information available from a static checking of LF types,
and it refined the resulting \hhf\ specifications
towards making their structure more closely resemble that of the LF
specifications they originated from.

The primary motivation for the work of Snow~\etal was a desire to use
Teyjus as a backend for an alternative implementation of logic
programming in Twelf.
However, it falls short of achieving this goal in two ways that we
address in this paper.
First, although it relates derivations from LF specifications to ones
from their translations, it does not make explicit the process of
extracting an LF ``result'' term from a successful \hhf~derivation; such
an extraction is necessary if Teyjus is to serve as a genuine,
invisible backend.
To close this gap, we describe an inverse translation and show that it
has the necessary properties to allow Twelf behavior to be emulated
through computations from $\lambda$Prolog programs.
Second, Snow~\etal dealt only with closed types, \ie, they did not
treat the idea of filling in missing parts of types in the course of
looking for an inhabitant.
To overcome this deficiency, we include meta-variables in
specifications and treat them in the back-and-forth translations as
well as in derivations; the last aspect, that is also the most
critical one in our analysis, requires us to build substitutions and
unification explicitly into our formalization of derivations.

The remainder of this paper is structured as follows.
Sections~\ref{sec:twelf} and \ref{sec:hohh} respectively present LF
and the \hhf~logic together with their computational interpretations.
Section~\ref{sec:translation} describes a
translation from LF specifications into \hhf~ones together with an
inverse translation for extracting solution terms from
\hhf~derivations.
We then propose an approach for developing a proof of correctness for this 
translation.
Section~\ref{sec:optimizations} improves the basic
translation and Section~\ref{sec:example} uses it to illustrate our
proposed approach to realizing logic programming in Twelf.
Section~\ref{sec:conclusion} concludes the paper.

  \section{Logic programming in LF}
\label{sec:twelf}

Three categories of expressions constitute LF: kinds, type families or
types which are classified by kinds, and objects which are classified
by types.
Below, $x$ denotes an object variable, $X$ an object
meta-variable, $c$ an object constant, and $a$ a type constant.
Letting $K$ range over kinds, $A$ and $B$ over types, and $M$ and $N$
over objects, the syntax of these expressions is given as follows:
\begin{tabbing}
  \qquad\=$A,B$\quad\=$::=$\quad\=\kill
  \>$K$ \> $::=$ \> $\type\ |\ \typedpi{x}{A}{K}$ \\
  \>$A,B$ \> $::=$ \> $a\ |\ \typedpi{x}{A}{B}\ |\ \app{A}{M} $ \\
  \>$M,N$ \> $::=$ \> $c\ |\ x\ |\ X\ |\ \typedlambda{x}{A}{M}\ |\ \app{M}{N}$
\end{tabbing}
Both $\Pi$ and $\lambda$ are binders which also assign types to the
(object) variables they bind over expressions.
Notice the dependency present in LF expressions: a bound object
variable may appear in a type family or kind.
In the remainder of this paper we use $U$ and $V$ ambiguously for types
and objects and $P$ similarly for types and kinds.
The shorthand $A\ra P$ is used for $\typedpi{x}{A}{P}$ if $P$ is a type
family or kind that is not dependent on the bound variable, i.e. if
$x$ does not appear free in $P$.
Terms differing only in bound variable names are identified.
We write $\subst{U}{M_1/x_1,\ldots,M_n/x_n}$ to denote the capture avoiding
substitution of $M_1,\ldots,M_n$ for the free occurrences of $x_1,...,x_n$
respectively in $U$.

LF kinds, types and objects are formed relative to a signature
$\Sigma$ that identifies constants together with their kinds or types.
In determining if an expression is well-formed, we additionally
need to consider contexts, denoted by $\Gamma$, that assign types to
variables.
The syntax for signatures and contexts is as follows:
\[\Sigma\quad ::=\quad \emptyctx
\ |\ \Sigma,\oftype{a}{K}\ |\ \Sigma,\oftype{c}{A} \qquad\qquad\qquad
\Gamma\quad ::=\quad \emptyctx \ |\ \Gamma,\oftype{x}{A} \]
In contrast to usual LF presentations, we have allowed
expressions to contain object meta-variables.
We assume an infinite supply of such variables for each type and that
an implicit meta-variable context $\Delta$ assigns types to these
variables.
These meta-variables act as placeholders, representing the part of an
expression one wishes to leave unspecified.

\begin{figure}[t]
\begin{center}

\AxiomC{$X:A\in\Delta$}
\RightLabel{\varext}
\UnaryInfC{$\lfprovesig{\Gamma}{\Sigma}{\oftype{X}{A}}$}
\DisplayProof

\vspace{10pt}

\begin{tabular}{ccc}
   \AxiomC{$\Sigma~\sig \quad \oftype{c}{A} \in \Sigma$}
   \RightLabel{\constobj}
   \UnaryInfC{$\lfprovesig{\Gamma}{\Sigma}{\oftype{c}{A^\beta}}$}
   \DisplayProof

&
\qquad\qquad
&
   \AxiomC{$\lfprovesig{\Gamma}{\Sigma}{\oftype{A}{\type}} \quad
     \lfprovesig{\Gamma,\oftype{x}{A}}{\Sigma}{\oftype{M}{B}}$}
   \RightLabel{\absobj}
   \UnaryInfC{$\lfprovesig{\Gamma}{\Sigma}
              {\oftype{(\typedlambda{x}{A}{M})}{(\typedpi{x}{A^\beta}{B})}}$}
   \DisplayProof

\end{tabular}

\vspace{10pt}

\begin{tabular}{ccc}
   \AxiomC{$\lfprovesig{}{\Sigma}{\Gamma\ \ctx}\quad\oftype{x}{A}\in \Gamma$}
   \RightLabel{\varobj}
   \UnaryInfC{$\lfprovesig{\Gamma}{\Sigma}{\oftype{x}{A^\beta}}$}
   \DisplayProof

&
\qquad\qquad
&

   \AxiomC{$\lfprovesig{\Gamma}{\Sigma}{\oftype{M}{\typedpi{x}{A}{B}}} \quad
     \lfprovesig{\Gamma}{\Sigma}{\oftype{N}{A}}$}
   \RightLabel{\appobj}
   \UnaryInfC{$\lfprovesig{\Gamma}{\Sigma}{\oftype{(\app{M}{N})}{(B[N/x])^\beta}}$}
   \DisplayProof
\end{tabular}
\end{center}
\vspace{-0.25cm}
\caption{Rules for typing LF objects}
  \label{fig:lf-rules}
\vspace{-0.5cm}
\end{figure}

Complementing the syntax rules, LF has typing rules that limit
the set of acceptable or well-formed expressions.
These rules define the following mutually recursive judgments with the
associated declarative content:
\begin{tabbing}
\qquad\=$\lfprovesig{\Gamma}{\Sigma}{\oftype{M}{A}}$\qquad\=\kill
\>$\Sigma~\sig$\>$\Sigma$ is a valid signature\\
\>$\lfprovesig{}{\Sigma}{\Gamma~\ctx}$\>$\Gamma$ is a valid context
relative to the (valid) signature $\Sigma$\\
\>$\lfprovesig{\Gamma}{\Sigma}{K~\kind}$\>$K$ is a valid kind in signature
$\Sigma$ and context $\Gamma$\\
\>$\lfprovesig{\Gamma}{\Sigma}{\oftype{A}{K}}$\>$A$ is a type of kind
$K$ in a signature $\Sigma$ and context $\Gamma$\\
\>$\lfprovesig{\Gamma}{\Sigma}{\oftype{M}{A}}$\>$M$ is an
object of type $A$ in signature $\Sigma$ and context $\Gamma$
\end{tabbing}
In our discussion of logic programming, we rely on a specific
knowledge of the rules for only the last of these judgments which we
present in Figure~\ref{fig:lf-rules}; an intuition for the other rules
should follow from the ones presented and their explicit presentation
can be found, \eg, in \cite{harper93jacm}.
By these rules we can see that if a well-formed expression contains a meta-
variable $X$ of type $A$, then replacing the occurrences of $X$ with a well-
formed object of type $A$ will produce an expression which is also well-formed.

The rules in Figure~\ref{fig:lf-rules} make use
of an equality notion for LF expressions that is based on
$\beta$-conversion, \ie, the reflexive and transitive closure of a
relation equating two expressions which differ only in that a
subexpression of the form $(\app{(\typedlambda{x}{A}{M})}{N})$  in one is
replaced by $\subst{M}{N/x}$ in the other.
We shall write $U^\beta$ for the $\beta$-normal form of an expression,
\ie, for an expression that is equal to $U$ and that does
not contain any subexpressions of the form
$(\app{(\typedlambda{x}{A}{M})}{N})$.
Such forms are not guaranteed to exist for all LF
expressions.
However, they do exist for well-formed LF expressions
\cite{harper93jacm}, a property that is ensured to hold for each
relevant LF expression by the premises of every rule whose conclusion
requires the $\beta$-normal form of that expression.

Equality for LF expressions also includes $\eta$-conversion, \ie, the
congruence generated by the relation that equates
$\typedlambda{x}{A}{(\app{M}{x})}$ and $M$ if $x$ does not appear free
in $M$.
The $\beta$-normal forms for the different categories of
expressions have the following structure
\begin{tabbing}
  \=\qquad\=\qquad\quad\ \=\kill
  \>\>$Kind$ \> $\typedpi{x_1}{A_1}{\ldots\typedpi{x_n}{A_n}{Type}}$\\
  \>\>$Type$ \> $\typedpi{y_1}{B_1}{\ldots\typedpi{y_n}{B_m}{a\ M_1\ \ldots\ M_n}}$\\
  \>\>$Object$ \> $\typedlambda{x_1}{A_1}{\ldots\typedlambda{x_n}{A_n}{u\ M_1\ \ldots\ M_n}}$
\end{tabbing}
where $u$ is an object constant or variable and where the subterms and
subtypes appearing in the expression recursively have the same form.
We refer to the part corresponding to $a\ M_1\ \ldots\ M_n$ in a
type in this form as its \emph{target} type and to
$B_1,\ldots,B_m$ as its \emph{argument} types.
Let $w$ be a variable or constant which appears in the well-formed
term $U$ and let the number of $\Pi$s that appear in the prefix of its
type or kind in beta normal form be $n$.
We say $w$ is {\it fully applied} if every occurrence of $w$ in $U$
has the form $\app{w}{M_1\ldots M_n}$.
A type of the form $\app{a}{M_1\ldots M_n}$ where $a$ is fully applied
is a {\it base type}.
We also say that $U$ is {\it canonical} if it is in normal form and every
occurrence of a variable or constant in it is fully applied.
It is a known fact that every well-formed LF expression is equal to
one in canonical form by virtue of
$\beta\eta$-conversion~\cite{harper93jacm}.
For the remainder of this paper we will assume all terms are in 
$\beta$-normal form.

A specification in LF comprises a signature that, as we have
seen, identifies a collection of object and type constants.
The Curry-Howard isomorphism \cite{howard80} allows types to be
interpreted dually as formulas.
The dependent nature of the LF type system allows type constants to
take objects as arguments.
Such constants then correspond to the names of predicates over
suitably typed objects.
Moreover, the same isomorphism allows object constants, which provide
a means for constructing expressions of particular types, to be viewed
as the names of parameterized rules for constructing proofs of the
relations represented by the types.

\begin{figure}[t]
\begin{verbatim}
nat : type.                   list : type.
z : nat.                      nil : list.
s : nat -> nat.               cons : nat -> list -> list.

append : list -> list -> list -> type.
app-nil : append nil L L.
app-cons : append L1 L2 L3 -> append (cons X L1) L2 (cons X L3).
\end{verbatim}
\vspace{-0.25cm}
\caption{A Twelf signature specifying lists and the append relation}
\label{fig:lf_app}
\vspace{-0.5cm}
\end{figure}

Figure~\ref{fig:lf_app} presents a concrete signature to illustrate
these ideas.
In showing this and other similar signatures, we use the Twelf syntax
for LF expressions.
In this syntax, $\typedpi{x}{A}{U}$ is written as $\{x:A\}\ U$ and
$\typedlambda{x}{A}{M}$ is written as $[x:A]\ M$.
Further, bindings and the corresponding type annotations on variables
are made implicit in situations where the types can be uniquely
inferred; the variables that are implicitly bound are denoted in
Prolog style by tokens that begin with uppercase letters.
The initial part of the signature in Figure~\ref{fig:lf_app}
defines type and object constants that provide a representation
of the natural numbers and lists of natural numbers.
The signature then identifies a type constant \verb+append+ that
takes three lists as arguments.
Under the viewpoint just explained, this constant can be interpreted
as a predicate that relates three lists.
Objects of this type can be constructed by using the constants
\verb+app-nil+ and \verb+app-cons+ that are also presented in the
signature.
Viewed differently, these constants name rules that can be used to
construct a proof of the append relation between three lists.
Notice that \verb+app-cons+ requires as an argument an object of
\verb+append+ type.
This object plays the role of a premise for the rule that
\verb+app-cons+ identifies.

The logic programming use of LF that underlies Twelf consists of
presenting a type $A$ in the setting of a signature $\Sigma$.
Such a type corresponds to the request to find an object $M$ such that
the judgment  $\lfprovesig{}{\Sigma}{M:A}$ is derivable.
Alternately, a query in Twelf can be seen as the desire to determine
the derivability of a formula, the inhabiting term that is found being
its proof.
The type that is presented as a query may also contain meta-variables,
denoted by tokens that begin with uppercase letters.
In this case, the request is to find substitutions for these variables
while simultaneously showing that the instance type is inhabited.

An example of a query relative to the signature in
Figure~\ref{fig:lf_app} is the following.
\begin{verbatim}
   append (cons z nil) nil L
\end{verbatim}
An answer to this query is the substitution \verb+(cons z nil)+
for \verb+L+, together with the object
\verb+(app-cons (cons z nil) nil (cons z nil) (app-nil nil))+
that inhabits that type. Another query in this setting is
\begin{verbatim}
  {x:nat} append (cons x nil) (cons z (cons x nil)) (L x).
\end{verbatim}
in which \verb+L+ is a ``higher-order'' meta-variable of type
\verb+nat -> list+. The substitution that would be computed by
Twelf for the variable \verb+L+ in this query is
\begin{verbatim}
   [y:nat] (cons y (cons z (cons y nil))),
\end{verbatim}
and the corresponding inhabitant or proof term is
\begin{verbatim}
   [y:nat] app-cons nil (cons z (cons y nil))
                        (cons z (cons y nil)) y
                        (app-nil (cons z (cons y nil)))
\end{verbatim}
Notice that the variable \verb+x+ that is explicitly bound in the
query has a {\it different} interpretation from the meta-variable
\verb+L+.
In particular, it receives a ``universal'' reading: the query
represents a request to find a value for \verb+L+ that yields an
inhabited type regardless of what the value of \verb+x+ is.

Although neither of our example queries exhibited this behavior, the
range of an answer substitution may itself contain variables and
there may be some residual constraints on these variables presented in
the form of a collection of equations between object expressions
called ``disagreement pairs.''
The interpretation of such an answer is that a complete solution can
be obtained from the provided substitution by instantiating the
remaining variables with closed object expressions that render
identical the two sides of each disagreement pair.

  \section{Logic programming based on  \hhf}
\label{sec:hohh}
\begin{figure}[t]
\begin{center}
\begin{tabular}{ccccc}
    \AxiomC{}
    \RightLabel{\topGoal}
    \UnaryInfC{$\sigsequent{\Xi}{\Gamma}{\top}$}
    \DisplayProof

&
\quad\quad

&
    \AxiomC{$\sigsequent{\Xi}{\Gamma \cup\{D\}}{G}$}
    \RightLabel{\impGoal}
    \UnaryInfC{$\sigsequent{\Xi}{\Gamma}{D \supset G}$}
    \DisplayProof

&
\quad\quad
&
    \AxiomC{$c \notin \Xi \quad \sigsequent{\Xi \cup
        \{c\}}{\Gamma}{G[c/x]}$}
    \RightLabel{\allGoal}
    \UnaryInfC{$\sigsequent{\Xi}{\Gamma}{\forallx{x}{G}}$}
    \DisplayProof

\end{tabular}

\vspace{10pt}

\begin{tabular}{c}
    \AxiomC{$\sigsequent{\Xi}{\Gamma}{\subst{G_1}{\myvec{t_1/x_1}}}\qquad \ldots \qquad
      \sigsequent{\Xi}{\Gamma}{\subst{G_n}{\myvec{t_1/x_1},\ldots,\myvec{t_n/x_n}}}$}
    \RightLabel{\bcGoal}
    \UnaryInfC{$\sigsequent{\Xi}{\Gamma}{A}$}
    \DisplayProof\\[7pt]
    where
    $\forallx{\myvec{x_1}}{(G_1\supset\ldots\supset\forallx{\myvec{x_n}}{(G_n\supset
        A')\ldots)}} \in \Gamma$, \\[3pt]
    $\myvec{t_1},\ldots,\myvec{t_n}$ are $\Xi$-terms and
    $\subst{A'}{\myvec{t_1/x_1},\ldots,\myvec{t_n/x_n}} = A$
\end{tabular}
\end{center}
\vspace{-0.25cm}
\caption{Derivation rules for the \hhf~logic}
\vspace{-0.5cm}
\label{fig:hhfrules}
\end{figure}

An alternative approach to specifying formal systems is to use a
logic in which relationships between terms are encoded in predicates.
The idea of animating a specification then corresponds to
constructing a proof for a given ``goal'' formula in the chosen logic.
To yield a sensible notion of computation, specifications must
also be able to convey information about how a search for a proof
should be conducted.
Towards this end, we use here the logic of higher-order
hereditary Harrop formulas, referred to in short as the
\hhf~logic.
This logic underlies the programming language
$\lambda$Prolog~\cite{nadathur88iclp}.

The \hhf~logic is based on Church's Simple Theory of
Types~\cite{church40}.
The expressions of this logic are those of a simply typed
$\lambda$-calculus (STLC).
Types are constructed from the atomic type $o$ for propositions and a
finite set of other atomic types by using the function type
constructor $\ra$.
We assume we have been given a set of variables and a set of constants,
each member of these sets being identified together with a type.
More complex terms are constructed from these atomic symbols
by using application and $\lambda$-abstraction in a way that respects
the constraints of typing.
As in LF, terms differing only in bound variable names are identified.
The notion of equality between terms is further enriched by $\beta$-
and $\eta$-conversion.
When we orient these rules and think of them as reductions, we are
assured in the simply typed setting of the existence of a
unique normal form for every well-formed term under these reductions.
Thus, equality between two terms becomes the same as the identity of
their normal forms.
For simplicity, in the remainder of this paper we will assume that all
terms have been converted to normal form.
We write $\subst{t}{s_1/x_1,\ldots,s_n/x_n}$ to denote the capture
avoiding substitution of the terms $s_1,\ldots,s_n$ for free
occurrences of $x_1,...,x_n$ in $t$.

Logic is introduced into this setting by identifying a sub-collection
of the set of constants as logical constants and giving them a
special meaning.
The logical constants that we shall use here are the following:
\begin{tabbing}
  \=\qquad\=\quad\ \=\qquad\quad\ \=\kill
  \>\>$\top$ \> of type \> $o$\\
  \>\>$\supset$ \> of type \> $o\ra o\ra o$\\
  \>\>$\Pi$ \> of type \> $(\tau\ra o)\ra o$ for each type $\tau$
\end{tabbing}
We intend $\top$ to denote the always true proposition and $\supset$,
which we will write in infix form, to denote implication.
The symbol $\Pi$ corresponds to the generalized universal quantifier:
the usual notation $\forallx{x}{F}$ for universal quantification
serves as a shorthand for $\Pi(\lambdax{x}{F})$.

\begin{figure}[t]
\begin{verbatim}
nat : type.           list : type.
z : nat.              nil : list.
s : nat -> nat.       cons : nat -> list -> list.
                      append : list -> list -> list -> o.
\end{verbatim}
{\tt
$\forall$L. append nil L L.\\
$\forall$X$\forall$L1$\forall$L2$\forall$L3. append L1 L2 L3 $\supset$
  append (cons X L1) L2 (cons X L3).
}
\caption{An \hhf~specification of lists and the append relation}
\vspace{-0.5cm}
\label{fig:lp_app}
\end{figure}

To construct a specification within the \hhf~logic, a user must
identify a collection of types and a further set of constants, called
non-logical constants, together with their types.
A collection of such associations forms a signature.
There is a proviso on the types of non-logical constants: their
argument types must not contain $o$.
Non-logical constants that have $o$ as their target or result type
correspond to predicate symbols.
If $c$ is such a constant with the type $\tau_1\ra\ldots\ra\tau_n\ra
o$ and $t_1,\ldots,t_n$ are terms of type $\tau_1,\ldots,\tau_n$,
respectively, then the term $(c~t_1\ldots\ t_n)$ of type $o$
constitutes an \ital{atomic formula}.
We shall use the syntax variable $A$ to denote such formulas.
More complex terms of type $o$ are constructed from atomic formulas
by using the logical constants.
Such terms are also referred to as {\it formulas}.

The \hhf~logic is based on two special classes of formulas identified
by the following syntax rules:
\[
G\quad ::=\quad\top\ |\ A\ |\ D \supset G\ |\ \forallx{x}{G}
\qquad\qquad D\quad ::=\quad A\ |\ G \supset
D\ |\ \forallx{x}{D} \]
We will refer to a $D$-formula also as a program clause.
Notice that, in elaborated form, such a formula has the structure
$\forallx{\myvec{x_1}}{(G_1 \supset \ldots \supset
    \forallx{\myvec{x_n}}{(G_n \supset A)\ldots)}}$; we write
$\forall \myvec{x_i}$ here to denote a sequence of universal quantifications.

The computational interpretation of the \hhf~logic consists of
thinking of a collection of $D$-formulas as a program
and a $G$-formula as a goal or query that is to be solved against a
given program $\mathcal{P}$ in the context of a given signature $\Xi$.
We represent the judgment that the query $G$ has a solution in such a
setting by the ``sequent'' $\sigsequent{\Xi}{\mathcal{P}}{G}$.
The rules for deriving such a judgment are shown in
Figure~\ref{fig:hhfrules}.
Using these rules to search for a derivation leads to a process
in which we first simplify a goal in a manner determined by the logical
constants that appear in it and then employ program clauses in a familiar
backchaining mode to solve the atomic goals that are produced.
A property of the \hhf~logic that should be noted is that both
the program and the signature can change in the course of a
computation.

We illustrate the use of these ideas in practice by considering, once
again, the encoding of lists of natural numbers and the append
relation on them.
Figure~\ref{fig:lp_app} provides both the signature and the program
clauses that are needed for this purpose.
This specification is similar to one that might be provided in Prolog,
except for the use of a curried notation for applications and the fact
that the language is now typed.
We ``execute'' these specifications by providing a goal formula.
As with Twelf, we will allow goal formulas to contain free or
meta-variables for which we intend instantiations to be found through
proof search.
A concrete example of such a goal relative to the specification in
Figure~\ref{fig:lp_app} is {\tt (append (cons z nil) nil L)}.
This goal is solvable with the substitution {\tt (cons z nil)} for \verb+L+.
Another example of a query in this setting is $\forallx{\mbox{\tt
    x}}{{\mbox{\tt (append (cons x nil) (cons z (cons x nil)) (L x))}}}$
and an answer to this goal is the substitution $\lambdax{\mbox{\tt
    y}}{\mbox{\tt (cons y (cons z (cons y nil)))}}$ for \verb+L+.

  \section{Translating Twelf specifications into predicate form}
\label{sec:translation}
We now turn to the task of animating Twelf specifications using a
$\lambda$Prolog implementation.
Towards this end, we describe a meaning preserving translation from LF
signatures into \hhf~specifications.
Our translation extends the one in \cite{snow10ppdp} by allowing for
meta-variables in LF expressions.
We also present an inverse translation for bringing
solutions back from $\lambda$Prolog to the Twelf setting.

\begin{figure}[t]
\begin{minipage}{0.5\textwidth}
\centering
\begin{align*}
   \phi(A) &:= \lfobject \ \text{when $A$ is a base type} \\
   \phi(\typedpi{x}{A}{P}) &:= \phi(A)\ra \phi(P)\\
   \phi(\type) &:= \lftype
\end{align*}
\end{minipage}
\begin{minipage}{0.5\textwidth}
\centering
\begin{align*}
    \encTerm{u} &:= u \\
    \encTerm{x} &:= x \\
    \encTerm{X} &:= X \\
    \encTerm{M_1\ M_2} &:= \encTerm{M_1}\ \encTerm{M_2}\\
    \encTerm{\typedlambda{x}{A}{M}} &:= \lambda^{\phi(A)} x. \encTerm{M}
\end{align*}
\end{minipage}
\caption{Flattening of types and encoding of terms}
\vspace{-0.5cm}
\label{fig:lossy_enc}
\end{figure}

\begin{figure}[t]
\begin{align*}
    \enc{\typedpi{x}{A}{B}} :=&\
      \lambda M.~
      \forall x.~ (\app{\enc{A}}{x}) \supset (\enc{B}~(\app{M}{x})) \\
    \enc{A} :=&\
      \lambda M.~
      \hastype{M}{\encTerm{A}}
      \ \text{where $A$ is a base type}
\end{align*}
\vspace{-0.5cm}
\caption{Encoding of LF types using the {\tt hastype} predicate}
\vspace{-0.5cm}
\label{fig:judgements}
\end{figure}

The first step in our translation is to map dependently typed lambda
expressions into simply typed ones.
We shall represent both types and objects in LF by STLC terms (which
are also \hhf~terms),
differentiating the two categories by using the (simple)
type $\lfobject$ for the encodings of LF objects and $\lftype$ for
those of LF types.
To play this out in detail, we first associate an \hhf~type with each
LF type and kind that is given by the $\phi(\cdot)$ mapping shown in
Figure~\ref{fig:lossy_enc}.
Then, corresponding to each object and type-level LF constant $u:P$,
we identify an \hhf~constant with the same name but with type
$\phi(P)$.
Finally, we transform LF objects and kinds into \hhf~terms using the
$\encTerm{\cdot}$ mapping in Figure~\ref{fig:lossy_enc}.
%

We would like to consider an inverse to the transformation that we
have described above.
We have some extra information available in constructing such an
inverse: the constants that appear in the \hhf~terms of interest have
their correlates which have been given specific types in the originating LF
signature.
Even so, the lossy nature of the translation makes the inversion
questionable.
There are two kinds of problems.
First, because (the chosen) simple typing is not sufficiently
constraining, we may have well-formed STLC terms for which there is no
corresponding LF expression.
As a concrete example, consider the following LF signature:
\begin{verbatim}
    i : type     j : type     a : i -> j     c : i
\end{verbatim}
In the encoding we will have the following two constants with
associated types:
\begin{verbatim}
    a : lf-obj -> lf-obj                c : lf-obj
\end{verbatim}
This means that we can construct the simply typed term
\verb+(a (a c))+ which cannot be the image of any LF expression that
is well-formed under the given signature.
The second problem is that when an \hhf~term involves an abstraction,
the choice of LF type to use for for the abstracted variable is
ambiguous.
As a concrete example, consider the \hhf~term $\lambdax{x}{x}$ that
has the type \verb+lf-obj -> lf-obj+.
This term could map to the LF objects \verb+[x:nat] x+ and
\verb+[x:list] x+, amongst many other choices.

Our solution to these problems is twofold.
First, we will assume that we know the type of the LF object that the
inversion is to produce; this information
will always be available when the \hhf~terms arise in the course of
simulating LF typing derivations using \hhf~derivations.
Second, we will define inversion as a partial function: when we use it
to calculate an LF expression from an answer substitution returned by
an \hhf~computation, we will have an additional obligation to
show that the inverse must exist.

\begin{figure}[t]
\begin{center}
\begin{tabular}{ccc}
  \AxiomC{$X:A\in\Delta$}
\RightLabel{\invVar}
\UnaryInfC{$\invup{X}{A}{\Theta}{X}$}
\DisplayProof
&
\qquad\quad
&
  \AxiomC{$\invdown{M}{B}{\Theta,x:A}{M'}$}
  \RightLabel{\invAbs}
  \UnaryInfC{$\invdown{\lambdax{x}{M}}{\typedpi{x}{A}{B}}{\Theta}{\typedlambda{x}{A}{M'}}$}
  \DisplayProof
\end{tabular}
\\
[10pt]
  \AxiomC{$\invup{M_1}{\typedpi{x}{B}{A}}{\Theta}{M_1'}$ \qquad $\invdown{M_2}{B}{\Theta}{M_2'}$}
  \RightLabel{\invApp}
  \UnaryInfC{$\invup{\app{M_1}{M_2}}{\subs{M_2'/x}{A}}{\Theta}{\app{M_1'}{M_2'}}$}
  \DisplayProof\\
[10pt]

\begin{tabular}{ccc}
  \AxiomC{$u:A\in\Theta$}
  \RightLabel{\invConst}
  \UnaryInfC{$\invup{u}{A}{\Theta}{u}$}
  \DisplayProof

&
\qquad\qquad

&
  \AxiomC{$\invup{M}{A}{\Theta}{M'}$}
  \RightLabel{\invSyn}
  \UnaryInfC{$\invdown{M}{A}{\Theta}{M'}$}
  \DisplayProof
\end{tabular}
\end{center}
\vspace{-0.5cm}
\caption{An inverse encoding}
\vspace{-0.5cm}
\label{fig:inv}
\end{figure}

The rules in Figure~\ref{fig:inv} define the inverse transformation.
The judgments $\invdown{t}{A}{\Theta}{M}$ and
$\invup{t}{A}{\Theta}{M}$ are to be derivable when $t$ is an \hhf~term
in $\beta$-normal form that inverts to the LF object $M$ that has type
$A$ in a setting where variables and constants are typed according to
$\Theta$.
The difference between the two judgments is that the first expects $A$
as an input whereas the second additionally synthesizes the type.
The process starts with checking against an LF type---this type will
be available from the original LF query---and it is easily shown that
if  $\invdown{t}{A}{\Sigma\cup\Gamma}{M}$, then
$\lfprovesig{\Gamma}{\Sigma}{M:A}$.
Notice that we will only ever check an abstraction term against an LF
type, ensuring that the type chosen for the bound variable will be
unique.
We say a substitution $\theta$ is invertible in a given context and
signature if each term in its range is invertible in that setting,
using the type associated with the domain variable by $\Delta$.

The translation of LF expressions into \hhf~terms loses all
relational information encoded by dependencies in types.
For example it transforms the constants encoding the append relation
in Figure~\ref{fig:lf_app} into the following \hhf~signature:
\begin{verbatim}
   append : lf-obj -> lf-obj -> lf-obj -> lf-type.
   app-nil : lf-obj -> lf-obj.
   app-cons : lf-obj -> lf-obj ->
              lf-obj -> lf-obj -> lf-obj -> lf-obj.
\end{verbatim}
It is no longer possible to construe this as a specification of the
append relation between lists.
To recover the lost information, we employ a second pass that uses
predicates to encode relational content.
This pass employs the \hhf~predicate $hastype$ with type
$\lfobject\ra\lftype\ra o$ and generates clauses that are such that
$\hastype{X}{T}$ is derivable from them exactly when $X$ is the
encoding of an LF term $M$ of a base LF type whose encoding is $T$.
More specifically, this pass processes each item of the form $U:P$ in the
LF signature and produces from it the clause $\enc{P}\ \encTerm{U}$
using the rules in Figure~\ref{fig:judgements} that define $\enc{\cdot}$.

To illustrate the second pass, when used with the signature in
Figure~\ref{fig:lf_app}, we see that it will produce the following
clauses:
\begin{tabbing}
\quad\=\qquad\qquad\=\quad\=\quad\=\verb+hastype +\=\kill
\>\verb+hastype z nat.+\\
\> $\forall$\verb+x.hastype x nat +$\supset$\verb+ hastype (s x) nat.+\\
\> \verb+hastype nil list.+\\
\> $\forall$\verb+x.(hastype x nat +$\supset$\\
\>\>$\forall$\verb+l.(hastype l list +$\supset$ \verb+hastype (cons x l) list)).+\\
\\
\> $\forall$\verb+l.hastype l list +$\supset$\verb+ hastype (app-nil l) list.+\\
\>$\forall$\verb+x.(hastype x nat +$\supset\forall$\verb+l1.(hastype l1 list +$\supset$\\
\>\>$\forall$\verb+l2.(hastype l2 list +$\supset\forall$\verb+l3.(hastype l3 list +$\supset$\\
\>\>\>$\forall$\verb+a.(hastype a (append l1 l2 l3)+$\supset$\\
\>\>\>\>\verb+hastype (app-cons x l1 l2 l3 a)+\\
\>\>\>\>\>\verb+(append (cons x l1) l2 (cons x l3))))))).+
\end{tabbing}
Contrasting these clauses with the ones of the $\lambda$Prolog program
in Figure~\ref{fig:lp_app}, we see that it is capable not only of
producing answers to \verb+append+ queries but also a ``proof-term''
that traces the derivation of such queries.

The correctness of our translation is captured by the following
theorem (whose proof is currently incomplete).
We had said earlier that when looking at terms that are produced by
\hhf~derivations from LF translations, we would have an assurance that
these terms are invertible.
This is a property that flows, in fact, from the structure of the
$hastype$ clauses: as a \hhf~derivation is constructed, all the
substitution terms that are generated are checked to be of the right
type using the $hastype$ predicate, and so we will not be able to
construct a term which is not invertible.

\begin{theorem}
Let $\Sigma$ be an LF signature and let $A$ be an LF type that
possibly contains meta-variables.
\begin{enumerate}
\item If Twelf solves the query $M:A$ with the ground answer
  substitution $\sigma$, then there  is an invertible answer
  substitution   $\theta$ for the   goal $\enc{A}\ \encTerm{M}$ wrt
  $\enc{\Sigma}$ such that the
  inverse $\theta'$ of $\theta$ generalizes $\sigma$ (i.e. there
  exists a $\sigma'$ such that $\sigma'\circ\theta'=\sigma$).

\item If $\theta$ is an invertible answer substitution for
  $\enc{A}\ \encTerm{M}$, then its inverse is an answer substitution
  for $M:A$.
\end{enumerate}
\end{theorem}
Our approach to proving this theorem is to consider the operational
semantics of the two systems and to show that derivations in each
system can be factored into sequences of steps that can be simulated
by the other system.
Moreover, this simulation  ensures the necessary relationships hold
between the answer substitutions that are gradually developed by the
derivations in the respective systems.

  \section{Optimizing the translation}
\label{sec:optimizations}
The translation presented in the preceding section does not lend
itself well to proof search because it generates a large amount of
redundant typing checking.
There are many instances when this redundancy can be recognized
by a direct analysis of a given Twelf specification: in particular, we
can use a structural analysis of an LF expression to
determine that a term being substituted for a variable must be of the
correct type and hence it is unnecessary to check this explicitly.
In this section we develop this idea and present an improved
translation.
We also discuss another optimization that reflect the types in the
Twelf signature more directly into types in \hhf.
The combination of these optimizations produce clauses that are
more compact and that resemble those that might be written in
$\lambda$Prolog directly.

\begin{figure}[t]
\centering
    \AxiomC{$\termUV{dom(\Gamma); \cdot; x}{A_i}$ for some $A_i$ in $\myvec{A}$}
    \RightLabel{\ruvappt}
    \UnaryInfC{$\formulaUV{\Gamma; x}{c \myvec{A}}$}
\DisplayProof
\\[3pt]
\begin{tabular}{ccc}
  \AxiomC{$\formulaUV{\Gamma, \oftype{y}{A}; x}{B}$}
  \RightLabel{\ruvpit}
  \UnaryInfC{$\formulaUV{\Gamma; x}{\typedpi{y}{A}{B}}$}
\DisplayProof
&
\qquad\qquad
&
  \AxiomC{$\formulaUV{\Gamma_1;x}{B}$}
  \AxiomC{$\formulaUV{\Gamma_1,\oftype{y}{B},\Gamma_2;y}{A}$}
  \RightLabel{\ruvctxt}
  \BinaryInfC{$\formulaUV{\Gamma_1,\oftype{y}{B},\Gamma_2;x}{A}$}
\DisplayProof
\end{tabular}
\\[3pt]
  \AxiomC{$y_i \in \delta$ for each $y_i$ in $\myvec{y}$ \quad each
    variable in $\myvec{y}$ is distinct}
  \RightLabel{\ruvinito}
  \UnaryInfC{$\termUV{\Delta; \delta; x}{\app{x}{\myvec{y}}}$}
\DisplayProof
\\[3pt]
\begin{tabular}{ccc}
  \AxiomC{$y \notin \Delta$ and $\termUV{\Delta; \delta; x}{M_i}$ for
    some $M_i$ in $\myvec{M}$}
  \RightLabel{\ruvappo}
  \UnaryInfC{$\termUV{\Delta; \delta; x}{\app{y}{\myvec{M}}}$}
\DisplayProof
&
\qquad\qquad
&
  \AxiomC{$\termUV{\Delta; \delta, y; x}{M}$}
  \RightLabel{\ruvabso}
  \UnaryInfC{$\termUV{\Delta; \delta; x}{\typedlambda{y}{A}{M}}$}
\DisplayProof
\end{tabular}
\caption{Strictly occurring variables in types and objects}
\vspace{-0.5cm}
  \label{fig:ruvs}
\end{figure}

We are interested in translating an LF type of the form
$\typedpi{x_1}{A_1}{\ldots\typedpi{x_n}{A_n}{B}}$
into an \hhf~clause that can be used to determine if a type $B'$ can
be viewed as an instance $\subst{B}{M_1/x_1,\ldots,M_n/x_n}$ of the
target type $B$.
This task also requires us to show that $M_1,\ldots,M_n$ are
inhabitants of the types $A_1,\ldots,A_n$; in the naive translation,
this job is done by the $hastype$ formulas pertaining to $x_i$ and
$A_i$ that appear in the body of the \hhf~clause produced for the overall
type.
However, a particular $x_i$ may occur in $B$ in a manner which already
makes it clear that the term $M_i$ which replaces it in any instance of
$B$ must possess such a property.
What we want to do, then, is characterize such occurrences of $x_i$ such
that we can avoid having to include an inhabitation check in the \hhf~clause.

We define a strictness condition for variable
occurrences and, hence, for variables that possesses this kind of
property.
By using this condition, we can simplify the translation of a type
into an \hhf~clause without losing accuracy.
In addition to efficiency, such a translation also produces a result
that bears a much closer resemblance to the LF type from which it
originates.

The critical idea behind this criterion is that the
path down to the occurrence of $x$ is \emph{rigid}, \ie, it cannot be
modified by substitution and $x$ is not applied to arguments in a way
that could change the structure of the expression substituted for it.
We know that the structure will be unchanged by application of arguments by
requiring the occurrence of $x$ to be applied only to distinct $\lambda$-bound
variables.
Thus we know that any term substituted for $x$ has the correct type without
needing to explicitly check it.
Specifically, we say that the bound variable $x_i$ occurs strictly in
the type $\typedpi{x_1}{A_1}{\ldots\typedpi{x_n}{A_n}{B}}$ if it is
  the case that
\vspace{-0.2cm}
\[\formulaUV{x_1:A_1,\ldots,x_{i-1}:A_{i-1};x_i}
            {\typedpi{x_{i+1}}{A_{i+1}}{\ldots\typedpi{x_n}{A_n}{B}}}\]
holds.
We have been able to extend the strictness condition as described
in~\cite{snow10ppdp} recursively while preserving its utility in recognizing
redundancy in type checking.
We consider occurrences of bound variables to be strict
in the overall type if they are strict in the types of other bound
variables that occur strictly in the target type.
The relation defined in Figure~\ref{fig:ruvs} formalizes this idea.

When $\formulaUV{\Gamma;x}{A}$ is derivable it means that the variable $x$
appears strictly in the type $A$ in the context $\Gamma$.
As we work down through the structure of a type we will eventually look at a
specific term $M$ and a derivation of $\termUV{\Delta;\delta;x}{M}$ means that
$x$ appears strictly in the term $M$.
Here, $\Delta$ and $\delta$ are both lists of variables where $\delta$ contains
the $\lambda$-bound variables currently in scope,
while $\Delta$ contains the $\Pi$-quantified variables collected while walking
through the type $A$.

\begin{figure}[t]
  \begin{minipage}{0.45\textwidth}
\centering
\begin{align*}
   \phi(a\ M_1\ldots M_n) &:= a\mbox{\rm -}type \\
   \phi(\typedpi{x}{A}{P}) &:= \phi(A)\ra \phi(P)\\
   \phi(\type) &:= \lftype
\end{align*}
\end{minipage}
\begin{minipage}{0.45\textwidth}
\centering
\begin{align*}
    \encTerm{u} &:= u \\
    \encTerm{x} &:= x \\
    \encTerm{X} &:= X \\
    \encTerm{M_1\ M_2} &:= \encTerm{M_1}\ \encTerm{M_2}\\
    \encTerm{\typedlambda{x}{A}{M}} &:= \lambda^{\phi(A)} x. \encTerm{M}
\end{align*}
\end{minipage}
  \centering
  \begin{align*}
    \encExtP{\typedpi{x}{A}{B}}{\Gamma} :=&\
      \begin{cases}
        \lambda M.~ \forall x.~ \top \supset \encExtP{B}{\Gamma, x}(\app{M}{x})
          & \text{if}\ \formulaUV{\Gamma; x}{B} \\
        \lambda M.~ \forall x.~
            \encExtN{A}(x) \supset \encExtP{B}{\Gamma, x}(\app{M}{x})
          & \text{otherwise}
      \end{cases} \\
    \encExtP{u\ \overrightarrow{N}}{\Gamma} :=&\ \lambda M.~
        \app{u}{\overrightarrow{\encTerm{N}}}\ M
    \\
    \encExtN{\typedpi{x}{A}{B}} :=&\
      \lambda M.~ \forall x.~
         \encExtP{A}{\cdot}(x) \supset \encExtN{B}(\app{M}{x}) \\
    \encExtN{u\ \overrightarrow{N}} :=&\ \lambda M.~
        \app{u}{\overrightarrow{\encTerm{N}}}\ M
  \end{align*}
\vspace{-0.5cm}
\caption{Optimized translation of Twelf signatures to $\lambda$Prolog programs}
  \label{fig:improved}
\vspace{-0.5cm}
\end{figure}

Another, more direct, optimization is to reflect the LF types into types in the
simply typed lambda calculus.
Along with this optimization we can also use specialized predicates, rather than
just {\tt hastype}.
For each LF type $u:K$ we will create a new atomic type \verb+u-type+ in
\hhf, as well as a new predicate \verb+u+ which has the type
$\phi(K)$ \verb+-> u-type -> o+.
We then use these to encode the signature in a more natural way.
See Figure~\ref{fig:improved} for the new translation.

There are now two modes in which translation operates, the negative,
$\encExtN{\cdot}$, which is essentially the same as before in that it
does not check for strictness of bound variables, and the positive,
$\encExtP{\cdot}{}$, which will only generate $hastype$ formulas for
variables which do not appear strictly.
We do this to insure that the eliminations occur in situations in
which it makes sense to think of the implication encoding an
inhabitation check.
We will write $\forallx{x}{\encExtP{B}{\Gamma,x}(\app{M}{x})}$
for $\forallx{x}{\top\supset\encExtP{B}{\Gamma,x}(\app{M}{x})}$
in future to simplify the generated signatures.
These optimizations not only clean up the generated signature, but they also
improve performance as we have limited the number of clauses which match the
head of any given goal formula.

  \section{An illustration of the translation approach}
\label{sec:example}
We illustrate the use of the ideas described in the earlier sections
by considering the \verb+append+ relation specified in Twelf by the
signature in Figure~\ref{fig:lf_app}.
The Twelf query that we shall consider is the following that we
previously saw in Section~\ref{sec:twelf}:
\begin{verbatim}
  {x:nat} append (cons x nil) (cons z (cons x nil)) (L x).
\end{verbatim}
This query asks for a substitution for \verb+L+ that yields an
inhabited type and an object that is a corresponding
inhabitant.

\begin{figure}[t]
{\tt
\begin{tabbing}
\qquad\=\kill
nat : nat-type -> o.\\
list : list-type -> o.\\
append : list-type -> list-type -> list-type -> append-type -> o.\\
\\
nat z.\\
$\forall$x. nat x $\supset$ nat (s x).\\
list nil.\\
$\forall$x.(nat x $\supset\forall$l. list l $\supset$ list (cons x l)).\\
\\
$\forall$l. append nil l l (app-cons l).\\
$\forall$x$\forall$l1$\forall$l2$\forall$l3$\forall$a. append l1 l2 l3 a
$\supset$\\
\>append (cons x l1) l2 (cons x l3) (app-cons x l1 l2 l3 a).
\end{tabbing}
}
\vspace{-0.25cm}
\caption{The Twelf specification of {\tt append} translated into
  $\lambda$Prolog}
\vspace{-0.5cm}
\label{fig:lp_sig}
\end{figure}

Applying the optimized translation to the signature in
Figure~\ref{fig:lf_app} yields the $\lambda$Prolog program shown in
Figure~\ref{fig:lp_sig}.
Further, the Twelf query of interest translates into the \hhf~goal
formula
\begin{tabbing}
\qquad\=\kill
\>$\forall$\verb+x. append (cons x nil) (cons z (cons x nil)) (L x) M.+
\end{tabbing}
The answer substitution for this goal in $\lambda$Prolog is
\begin{verbatim}
  L = y\ cons y (cons z (cons y nil)),
  M = y\ app-cons nil (cons z (cons y nil))
                  (cons z (cons y nil)) y
                  (app-nil (cons z (cons y nil)))
\end{verbatim}
Applying the inverse translation described in
Section~\ref{sec:translation} to this answer substitution yields the
value for \verb+L+ and the proof term for the Twelf query that we saw
in Section~\ref{sec:twelf}.

  \section{Conclusion}
\label{sec:conclusion}
We have considered in this work an approach to implementing the logic
programming treatment of LF specifications that is embodied in Twelf
by using the Teyjus implementation of $\lambda$Prolog as a backend. 
Central to such an implementation is a meaning-preserving translation
of Twelf specifications into  $\lambda$Prolog programs.
The basic structure of such a translation has previously been
described by Snow~\etal \cite{snow10ppdp}.
Built into that translation is an optimization which takes advantage of
statically available type information, quantified through a notion of
strictness.
In this work we have refined the notion of strictness to potentially
enhance the usefulness of this optimization. 

To actually use this approach in an implementation of Twelf, it is
necessary to also provide a way of translating solutions found by
Teyjus into LF terms that constitute answers to the query in LF
syntax. 
Towards this end, we have presented an inverse encoding which
describes how to map \hhf~terms back to LF objects in the context of
the original Twelf specification.

The work by Snow \etal deals only with terms which are closed, and so there had 
been no treatment for meta-variables which may appear in LF expressions.
In order to capture the full scope of logic programming in Twelf, we extended 
the usual presentation of LF to allow for meta-variables in terms, and we 
provided a treatment for such variables in both the derivations and the 
translation.
Although the proof showing the correctness of this translation is still 
incomplete, we have discussed an approach to developing such a proof that is 
based on relating the operational semantics of the two systems.

  \section*{Acknowledgements}

This work has been partially supported by the NSF Grant CCF-0917140.
Opinions, findings, and conclusions or recommendations expressed in this paper
are those of the authors and do not necessarily reflect the views of the
National Science Foundation.

  \bibliographystyle{plain}
  \bibliography{master}

\end{document}